\documentclass[fleqn,10pt]{wlscirep}
\usepackage[utf8]{inputenc}
\usepackage[T1]{fontenc}
\title{A Data-Driven Approach to Violin Making }
\author[1,*]{Sebastian Gonzalez}
\author[1]{Davide Salvi}
\author[2]{Daniel Baeza}
\author[1]{Fabio Antonacci}
\author[1]{Augusto Sarti}
\affil[1]{Musical Acoustics Lab at the Violin Museum of Cremona, DEIB - Politecnico di Milano, Cremona Campus, Italy}

\affil[2]{Department of Electrical Engineering, Faculty of Physical and Mathematical Sciences, University of Chile, Chile}

\affil[*]{tsuresuregusa@gmail.com}

\begin{abstract}
Of all the characteristics of a violin, those that concern its shape are probably the most important ones, as the violin maker has complete control over them. Contemporary violin making, however, is still based more on tradition than understanding, and a definitive scientific study of the specific relations that exist between shape and vibrational properties is yet to come and sorely missed. In this article, using standard statistical learning tools, we show that the modal frequencies of violin tops can, in fact, be predicted from geometric parameters, and that artificial intelligence can be successfully applied to traditional violin making. We also study how modal frequencies vary with the thicknesses of the plate (a process often referred to as {\em plate tuning}) and discuss the complexity of this dependency. Finally, we propose a predictive tool for plate tuning, which takes into account material and geometric parameters. 

\end{abstract}
\begin{document}

\flushbottom
\maketitle

\thispagestyle{empty}

\section*{Introduction}

%If there was a reason behind the particular shape, it has been long lost in history. A more probable explanation was just a random evolution of the shape that created certain canonical models that sounded `great'
The violin made its first appearance in northern Italy in the early sixteenth century, and for nearly two centuries its shape kept gradually evolving until it reached a point of relative stability during the so-called ``Cremonese period". The city of Cremona, in fact, was already teaming with numerous luthiers, and the experimentation on string instruments was constantly in full throttle. At the beginning of the 18th century, Cremona was home to the most celebrated luthiers of all times, such as Antonio Stradivari and Giuseppe Guarneri ``del Ges\`u'', and through them this tradition of experimentation gave birth to some of the finest instruments ever made \cite{nia2015evolution,tai2018acoustic}. Nowadays, violin makers tend to follow a ``differential'' approach, and apply idiosyncratic variations to violin models of celebrated luthiers of that period.
To the best of our knowledge, there is no clear explanation as to which violin shape should be preferred to which, only anecdotal evidence of some shapes sounding `better' than others. 
In this article we try to shed light on this problem through a rather unconventional approach. Inspired by the evolution of the violin's shape along centuries, we set out to simulate thousands of violin tops and apply methods of machine intelligence to discover and understand the relations between shape and vibration. We show that a simple Neural Network (NN) can learn a great deal on how a certain geometry `sounds', i.e. what their eigenfrequencies are; and can be used for predicting results that otherwise would only be offered by Finite Element Method (FEM) simulations, the {\em de-facto} standard of simulation in violin research for the past four decades \cite{molin1984fem, molin1988parameters, tinnsten2002numerical,woodhouse2014acoustics, gough2015violin,torres2020exploring,chatziioannou2019reconstruction}. 

Machine intelligence has been successfully applied to physical systems of all sorts \cite{Carleo2019}, including spin phase transitions \cite{carrasquilla2017machine,van2017learning,shiina2020machine}; quantum topological transitions \cite{ming2019quantum}; and even physical problems as simple as the pendulum \cite{Iten2020}. In computational acoustics, NN's have been employed in a wide range of tasks\cite{bianco2019machine}, including the localisation of acoustic sources~\cite{chakrabarty2017broadband}; nearfield holography~\cite{olivieri2020nahcnn}; and acoustic scene classification~\cite{valenti2016dcase}. To the best of our knowledge, however, AI has not yet been applied to the problem of eigenfrequencies of plates, let alone the prediction of the acoustic behaviour of violin tops. 

\begin{figure}[ht]
\centering
\includegraphics[width=6in]{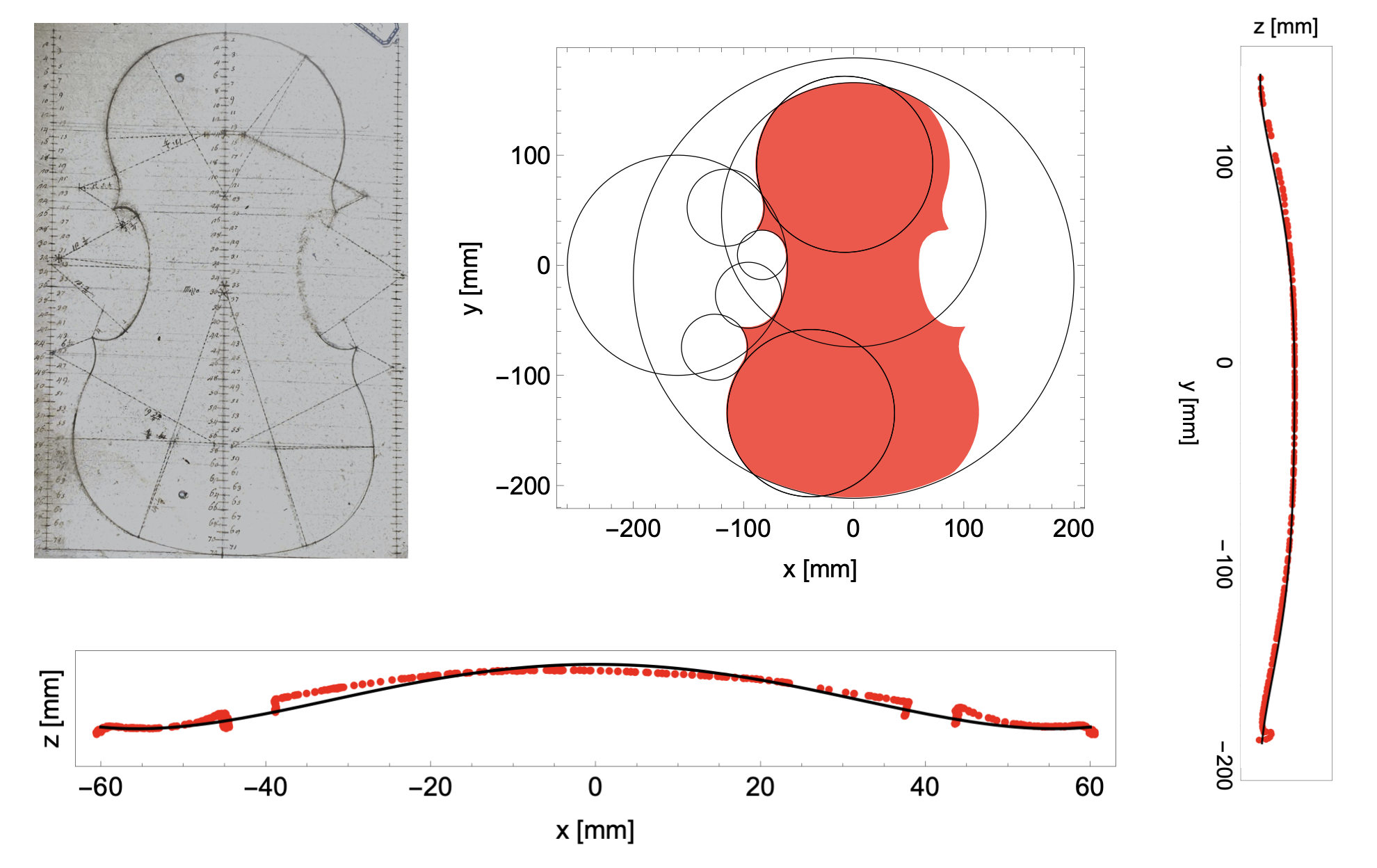}
\put(-435,255) { \bf a)}
\put(-280,255) { \bf b)}
\put(-70,255) { \bf c)}
\put(-420,70) { \bf d)}
%\\
%\includegraphics[width=6in]{arching0.pdf}

\caption{(a) A drawing made by Antonius Stradivarius showing the outline as a series of connected arcs of circles, preserved in (and courtesy of) the Violin Museum of Cremona, Italy\cite{cacciatori2016antonio}. (b) The 9 circles used for generating our violin outlines. (c) Fitting of a 6th order polynomial (black solid line) to the longitudinal arching (red points) of the celebrated ``Messiah" violin, made by Antonius Stradivarius in 1716 (part of the fingerboard is visible in the upper right corner). (d) Transversal arching profile measured at the centre (in red), obtained from the 3D scan of the ``Messiah"; arching of the 4th-order polynomial used (in black), see main text. }\label{fig.circles} 
\end{figure}
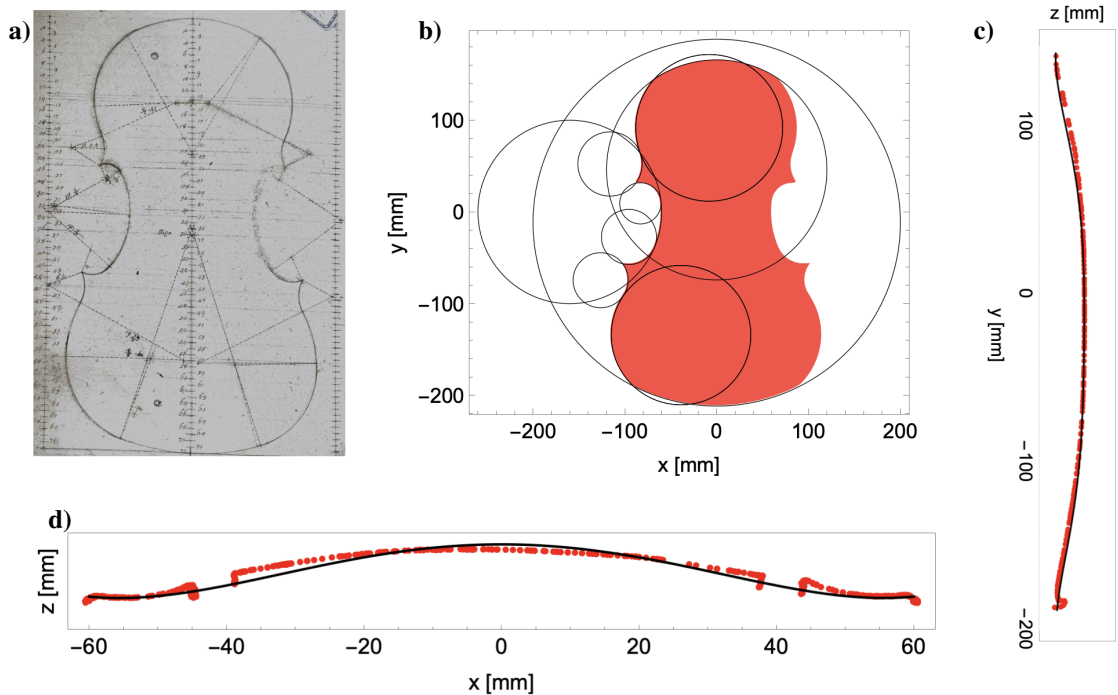

It is worth noticing that the eigenfrequencies (also referred to as modal frequencies) of the free plate are not immediately related to the acoustic properties of the complete instrument. They are, however, considered by violin makers as the parameters that drive the choices during the construction of the instrument. We proceed by first defining a parametric procedure to construct {\em in silico} the outline of the violin based on the drawings of Antonio Stradivari. This parametrisation allows us to create an arbitrary large dataset of violin shapes, which can then be used for answering the question as to whether AI can be used for predicting the eigenfrequencies. The answer, as we will see, is affirmative and we proceed to explain the way in which the prediction is done and study the correlation between eigenfrequencies and geometry of the violin. Finally, we use Principal Component Analysis (PCA) to show that the prediction is independent of the adopted parametrisation.  Although there is still a great deal of work to do to predict the violin's timbral features its eigenfrequencies, we consider this an important first step in that direction.  
\section*{Results}

\subsection*{Outline generation}
Our parametric construction of the outline of a violin top plate starts from a specific drawing of Antonius Stradivarius (Fig.~\ref{fig.circles}(a)), whose original is preserved in the Violin Museum of Cremona, Italy, where our laboratory is located. This drawing, in fact, suggests a parametric representation of the shape of the top plate. In our interpretation, a total of 9 arcs of circumference is combined together to form the different portions of the outline, as shown in Figure \ref{fig.circles}(b). Each one of these arcs is defined by the position of its centre, its radius and the two ``aperture" angles; and their connection must obey continuity constraints (each arc section begins where another one ends). This procedure ends up requiring 20 free parameters, which are described in the Section {\em Supplementary Material 1}. 
Through these parameters we control the outline of the violin, while the appropriate arching and thickness grading of the top plate can be independently controlled. For details on the data set creation, see \ref{methods}.

In order to confirm that the generated violin models are ``reasonable'', we asked a number of Cremonese luthiers to inspect the data set and judge if the resulting violin shapes are within the limits of what would be considered as ``canonical''. Interestingly enough, not only did they validate the dataset, but they also recognised the style of specific violin makers while browsing through the generated shapes: from more delicate Amati-like shapes to others that were more reminiscent of very specific ones such as the Stradivarius ``Hellier''.

\subsection*{Neural network prediction} 
\begin{figure*}[ht]
\centering
\includegraphics[width=2.7in]{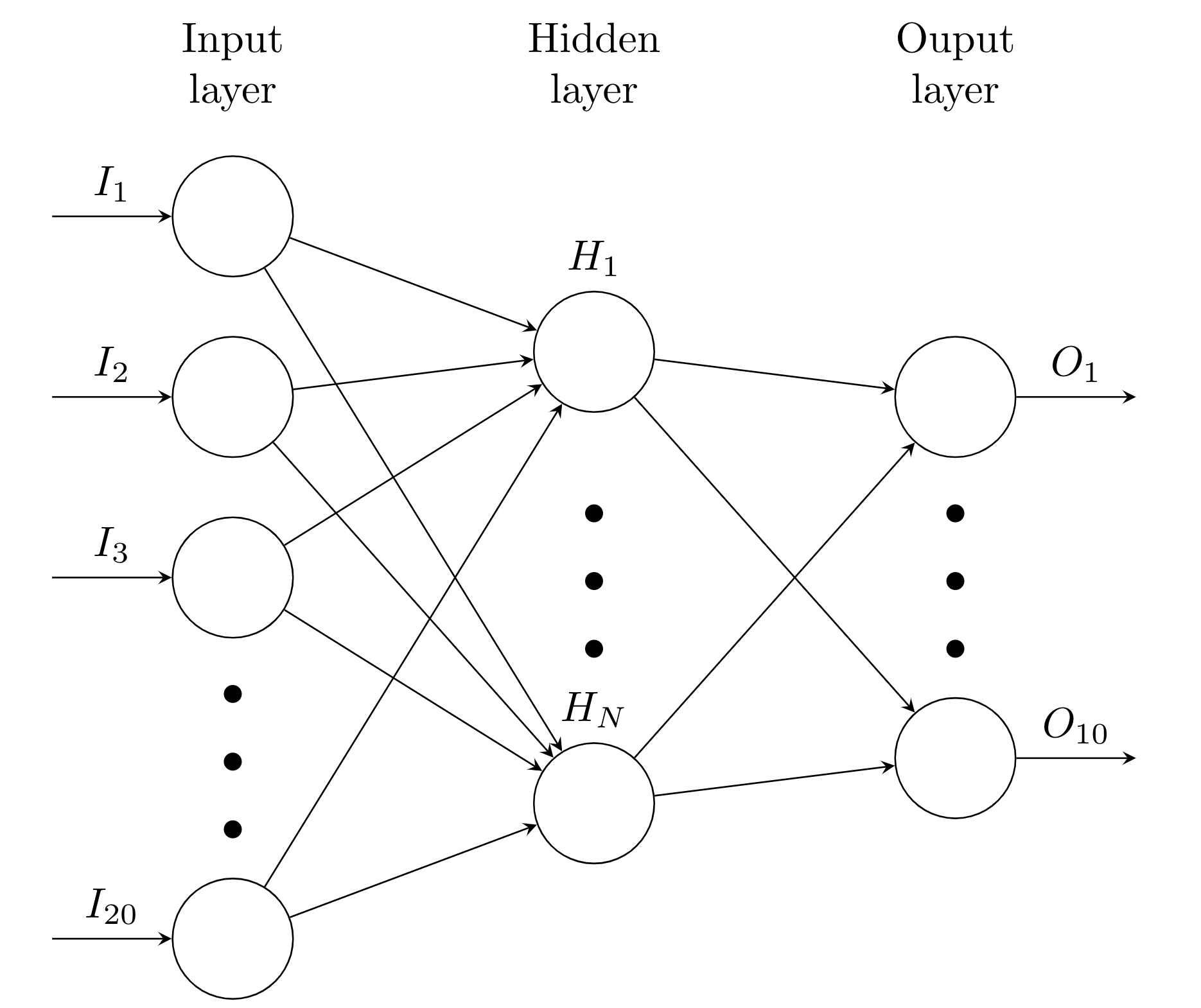}
 \put(-185,145) { \bf a)}
    \put(30,145) { \bf b)}
\hspace{.25in}
\includegraphics[width=3.in]{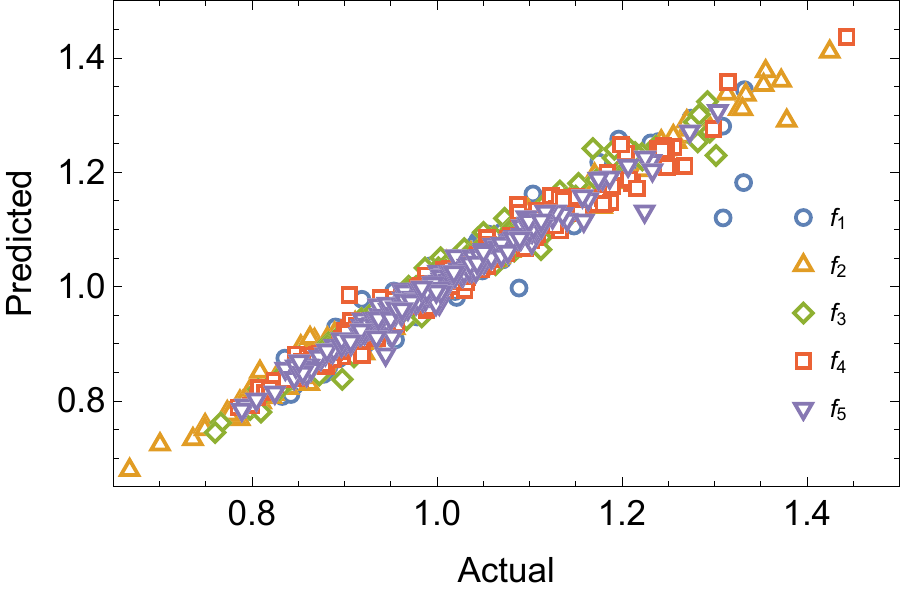}
\caption{a) Architecture used for prediction. b) Predicted versus actual values for the first five eigenfrequencies in the test set for a network with $N=7$. The frequencies are scaled by the average actual values for each mode in order to be able to compare different frequency values in the same plot. The prediction turns out to produce $R^2 = 0.977$.}\label{fig.predict}
\end{figure*}

In order to test the capability of a Neural Network to predict the eigenfrequencies of a plate for different outlines, we set up a simple architecture based on a hidden dense layer of $N$ neurons with sigmoid activation function, connected to a linear output layer, as shown in Fig.~\ref{fig.predict}(a). 
The inputs were the 20 parameters that define the outline; while the outputs were the first ten eigenfrequencies of the resulting top plate. We ran a total of 1750 simulations,
1500 of which we used for the training, and the remaining 250 we used for testing purposes.

\begin{figure}[ht]
\centering
\includegraphics[width=3.in]{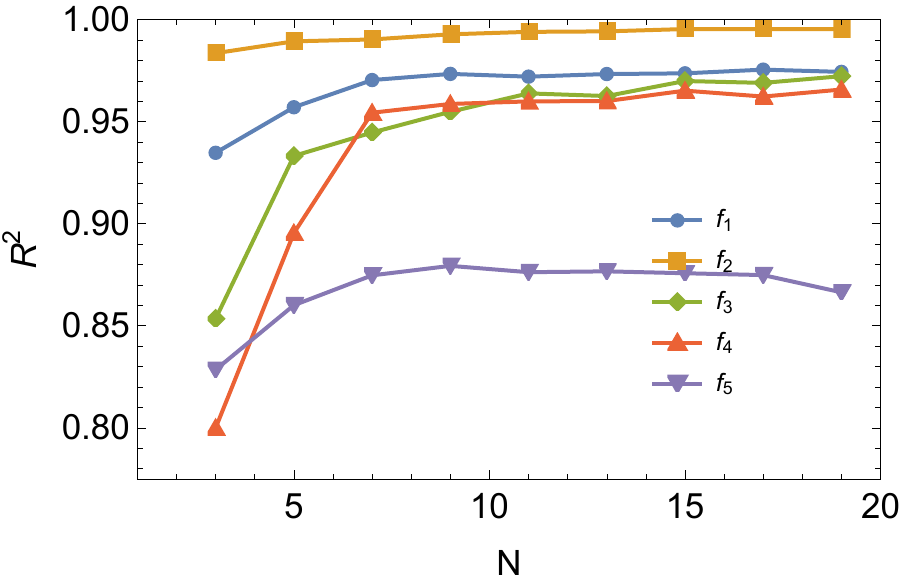}
 \put(-195,145) { \bf a)}
    \put(20,145) { \bf b)}
\hspace{.25in}
\includegraphics[width=3.in]{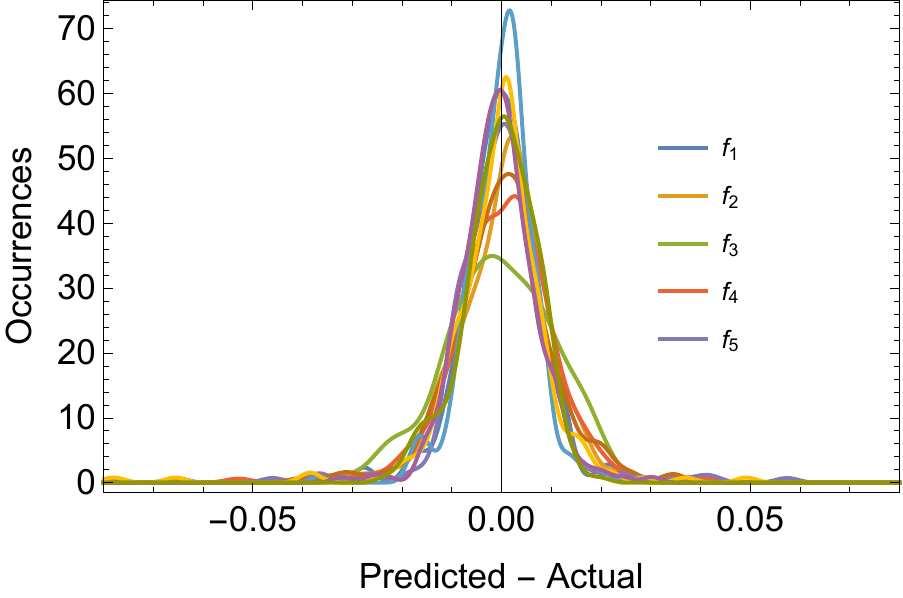}
\caption{(a) Individual values of the $R^2$ (predicted vs. actual) of the first five eigenfrequencies $f_{1,...,5}$ of the violin top plates. 
As we can see, from $N=15$ the network begins overfitting $f_5$ in the training set, as the error in the test set start growing. Notice that, with fewer than 7 neurons, the network is unable to offer a correct prediction. (b) Histograms of the difference between the eigenfrequency predicted by the neural network with N=7; and the Comsol\texttrademark\; result for the first five eigenfrequencies.}\label{fig.4}
\end{figure}

Figure \ref{fig.predict}(b) shows a comparison between the eigenfrequencies $f_{1,...,5}$ that are obtained through simulations and those predicted with the network with $N=7$. Of course, the evaluation is performed on a test set  made of violin tops that were not used in the training phase. The results are, all in all, remarkably accurate, though $f_5$ appears to produce a larger number of outliers. We also studied the accuracy of the network for a varying number of neurons in the hidden layer. Fig.~\ref{fig.4}(a) shows the $R^2$ of the simulated-vs-predicted values in the test set for all the frequencies, as a function of the number of hidden layers $N$. We immediately notice that the network offers good predictions of the eigenfrequencies and plateaus starting from $N=7$. From about $N=19$ on, the network begins overfitting the training set, and the error in the test set starts increasing. Interestingly enough, $f_5$ is consistently the hardest eigenfrequency to predict. If we compute the error for $N=50$ and $N=100$, the fit is indeed slightly worse than that of the other eigenfrequencies. Fig.~\ref{fig.4}(b) shows the distribution of the differences between prediction and simulation for a network with $N=7$, where the outliers in the prediction of $f_5$ are visible.

\subsection*{\label{sec:5}From black-box to white-box}
In order to assess how well the neural network was performing in its prediction task, we used ``feature importance analysis'', which is well-established in the context of machine learning. We computed, in fact, the permutation feature importance for the 20 geometrical parameters of the outline. We then studied how to use Principal Component Analysis (PCA) representations of the outline, in order to predict the eigenfrequencies. Finally, we studied the impact of the thickness profile and the material parameters on the frequencies and the outline. 

\subsubsection*{Permutation Feature Importance}
The {\em permutation feature importance} is defined as the reduction in the prediction accuracy that occurs when randomly shuffling a single feature value \cite{breiman2001random}. We used the model that was trained as described in the previous Section and we observed how the model score changes while permuting the input parameter $i$ at random for $n=10$ realisations. The metric that we use for assessing the accuracy is $s_{i,n} = R^2$, the coefficient of determination of actual vs predicted values for the eigenfrequencies, averaged over the $n$ realisations and for the first 5 eigenfrequencies of the top plate. The importance is defined and computed as
\begin{equation}
    I_i = s - \frac{1}{n}\sum_n s_{i, n}~,
\end{equation}
where $s=0.965$ is the value of $R^2$ of the test set with no permutations.

Figure \ref{fig.paramsCorr}(a) shows the results of the permutation feature importance for the outline prediction, in decreasing order.  The effect of each parameter in the outline can be seen in Fig. \ref{fig.paramsVar}. Here a 5\% variation is applied to selected parameters and the corresponding effect on the outline is shown. Notice that parameters 1 and 2 control the width of the violin, whereas parameters 14 and 16 control the lower bout size. Parameter 20, instead, controls the size of the upper bout. As previously mentioned, only the first 7 parameters carry a relevant amount of information of the frequencies. This suggests us that the neural network selects these parameters from the input and combines them in the hidden and output layers to obtain the eigenfrequency.

\subsubsection*{PCA prediction}
% Instead of working with the parameters of the model, we now interpret the outline as a collection of points. This is because when designing a violin, luthiers do not usually, as far as we know, use parametric modelling, but rather proceed according to their own rules and methods. In either case the outline of the violin is well defined. If we show that the neural network trained on the parametric model data can be used to predict also the eigenfrequencies (via the PCA decomposition), it is possible to envision its use in the workshop practice. 

\begin{figure}
    \centering
    \includegraphics[width=3in]{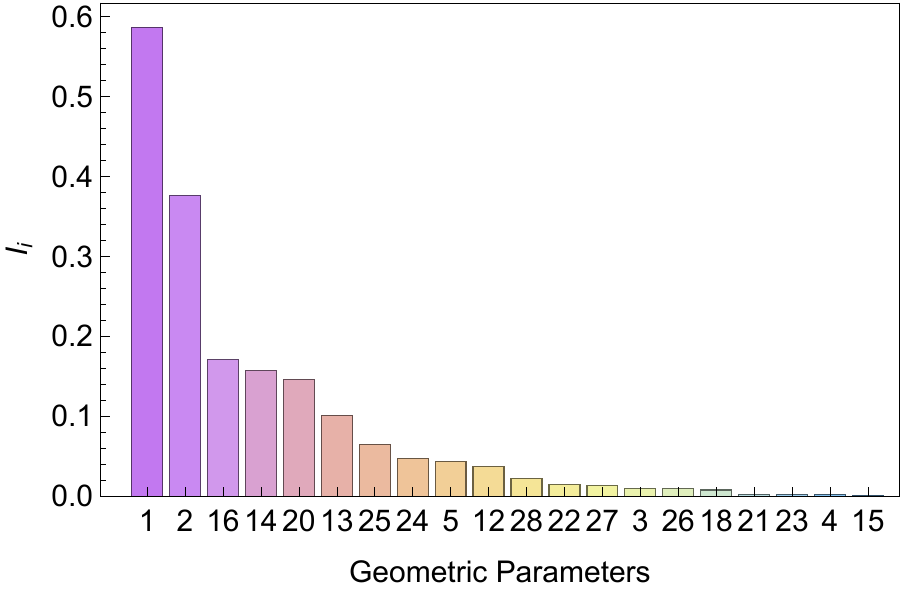}
    \put(-215,145) { \bf a)}
    \put(15,145) { \bf b)}
    \hspace{.25in}
    \includegraphics[width=3in]{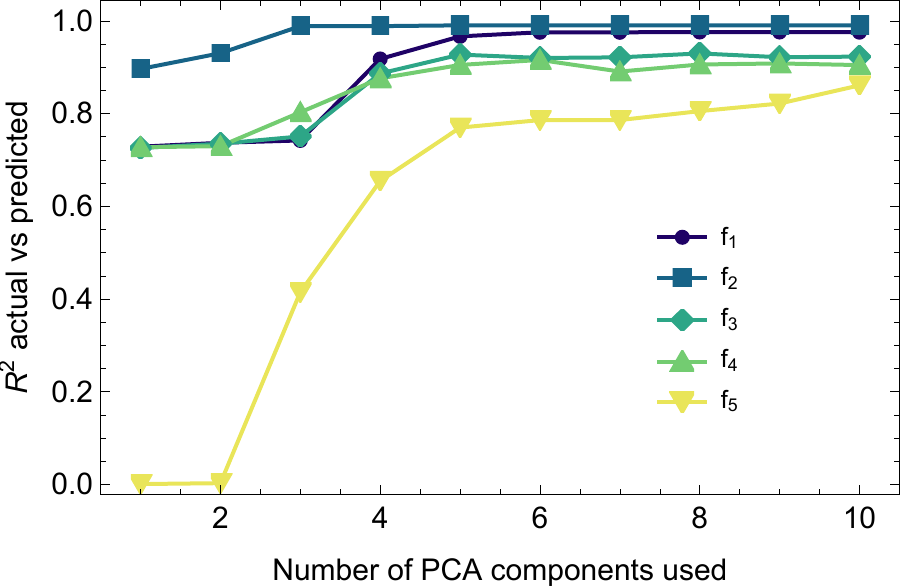}
    \caption{(a) Feature importance measured trough $I_i$. The geometric parameters are here sorted in decreasing order of importance. As we can see, only few parameters carry a significant amount of information about the eigenfrequencies. Interestingly, the parameters that correlate with frequency 5 are not the same that correlate the remaining frequencies. (b) Accuracy of the prediction model based on the PCA components of the outline instead of the geometric parameters, each line corresponds to a different mode frequency.}\label{fig.paramsCorr}
\end{figure}

\begin{figure}
    \centering
    \includegraphics[width=6in]{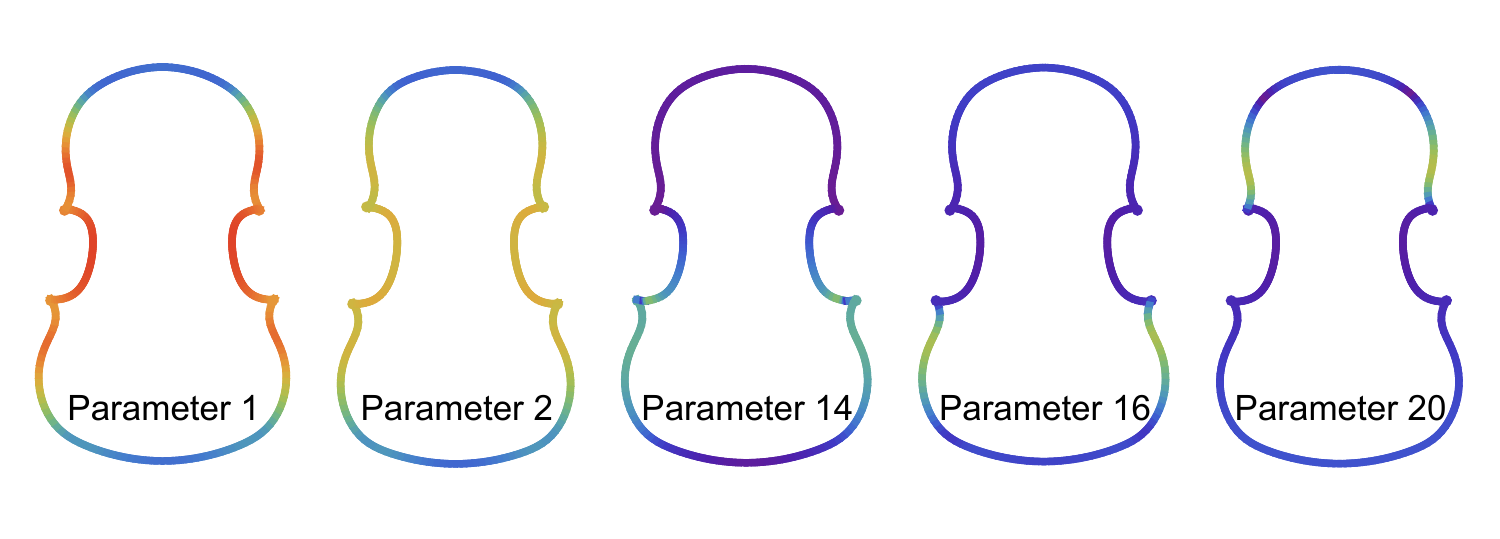}
    \includegraphics[width=.6in]{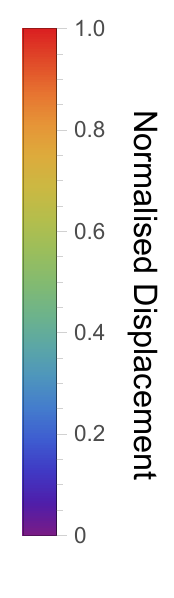}
    \caption{Outline change for a 5\% variation of the  for the first 5 parameters of the model, ordered by relevance from Fig.~\ref{fig.paramsCorr}. The colour code represents the displacement of each point of the outline, normalised by the maximum displacement, from the average violin. }
    \label{fig.paramsVar}
\end{figure}

\begin{figure}
    \centering
    \includegraphics[height=2in]{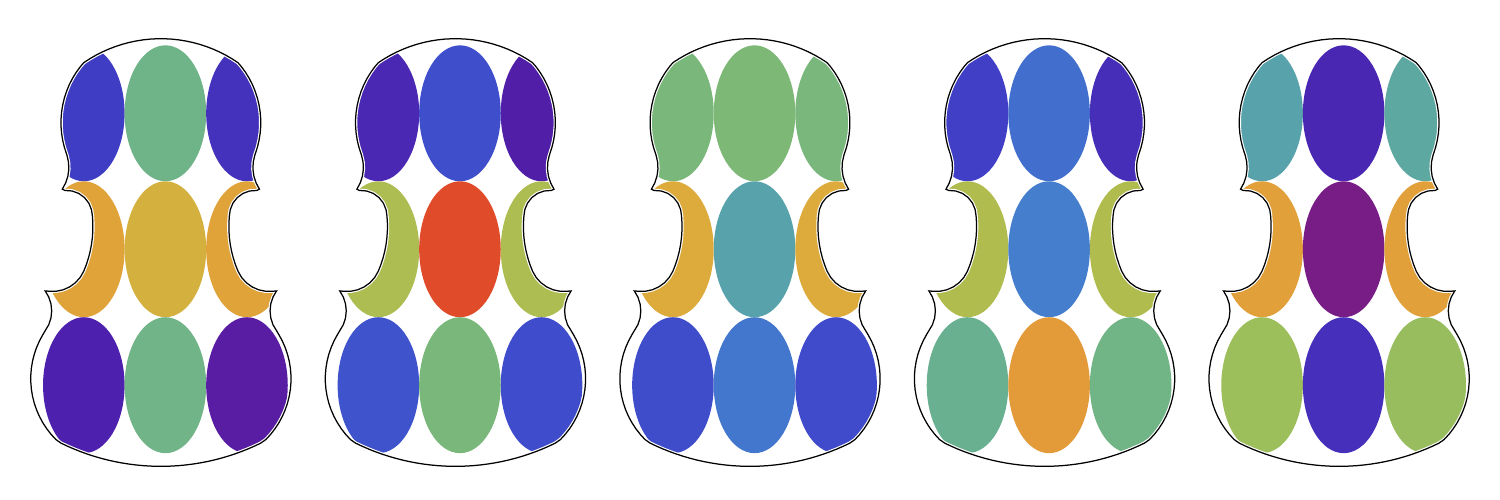}
    \put(-390,145) {  $f_1$}
    \put(-310,145) { $f_2$}
    \put(-225,145) { $f_3$}
    \put(-140,145) { $f_4$}
    \put(-55,145) { $f_5$}
    \includegraphics[height=2in]{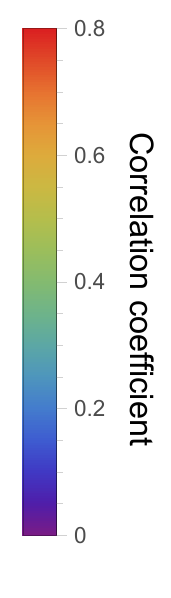}
        \caption{Correlation between thickness and  eigenfrequencies for modes one to five. Data set created varying only the thickness profile and keeping outline and material parameters constant. For the case of varying all the parameters at the same time the correlation values go down to a max of 0.5 but the spatial structure is conserved.}
    \label{fig.thicknessCorr}
\end{figure}
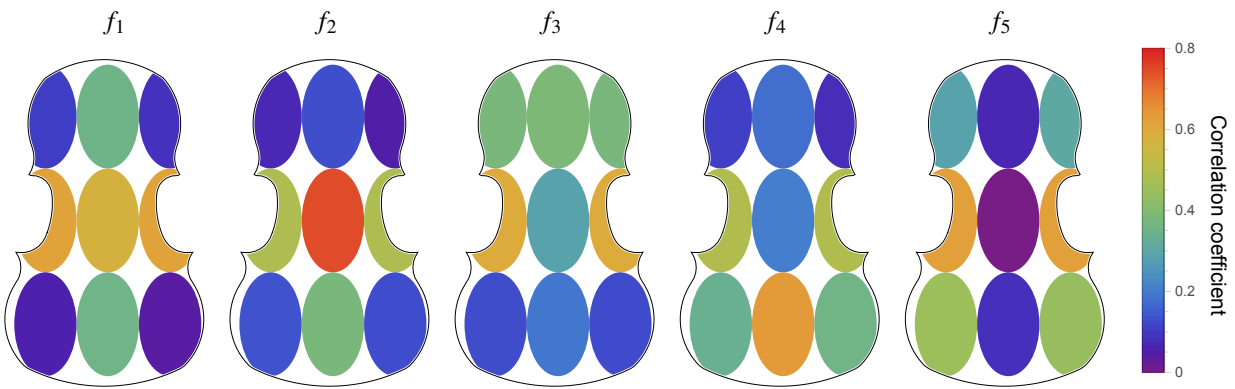

In order to study the relation between the outline and the eigenfrequencies, we first compute the PCA of the points of the outline. This is done by discretising the outline in 720 equispaced points. We concatenate the points of the outline and
rearrange into a single vector ${x_1, y_1,...,x_n,y_n}$ of length 1440. We compute the PCA over the entire set of 1750 violin top plates in the dataset. Already the first 10 PCA vectors are able to account for $98.8\%$ of the variance of the set. 

Figure \ref{fig.paramsCorr}(b) shows $R^2$ as a function of the number of PCA components of the outline used. This is akin to a coordinate transformation between two `coordinate systems'. It does not really matter to describe a violin in terms of circles or the outline, in the same way there is no difference writing equations in polar or Cartesian coordinates. Instead of predicting with the whole 20 parameters, we use the first 10  PCA components of the outline. This time we use a simple linear regression instead of a neural network. Through linear regression the prediction of the first 5 eigenfrequencies turns out to be very accurate, and the accuracy grows as more PCA components are added. The accuracy, however, is not uniform throughout the frequencies. For example, $f_2$ is extremely well predicted just with the first PCA component and, from the third PCA component on, not much additional information is gathered on that mode. Conversely, the first two PCA components do not contribute much to the prediction of $f_5$, and we need to go beyond such components to gain some knowledge on it. It is worth underlining that the accuracy achieved with the PCA of the outline is as good as the best prediction done with the full set of parameters despite using only half the degrees of freedom. This suggests that the predictive power of the violin's vibrational response rests with the geometry of the violin, rather than our particular parametrisation of choice.

\subsubsection*{Thickness profile}
Let us now focus on how the thickness profile influences the eigenfrequencies. The thickness profile is simply defined by 9 parameters, describing the thickness of the plate in 9 regions, as defined in \cite{gonzalez2020a}. The analysis is conducted on 1000 top plates, generated by varying the thickness in each one of the nine regions according to a Gaussian distribution with a $10\%$ spread. The correlation matrix between the first five eigenfrequencies and the thickness regions is shown in Fig.~\ref{fig.thicknessCorr}.
We immediately notice that there seems to be no single region in charge of one particular mode: any profile changes seem to simultaneously affect multiple frequencies, which debunks the widespread belief that certain modes can be controlled by removing material in certain areas (in particular the nodal lines) of the top plate. Furthermore,  $f_5$ appears to depend again on the lower bout, in this case its thickness. This is consistent with the fact that $f_5$ is most sensitive to changes in the parameters 14 and 16 of the outline, which are exactly those that control the width of the lower bout. 

\subsubsection*{Full parameter variation}
Let us finally consider the impact on the frequency response of varying the three different sets of parameter {\em individually} and {\em simultaneously}. 
We first created a data set where only the parameters of the material (density, stiffness, Young and shear moduli) changed according to a Gaussian distribution (for the actual formulas, see Table \ref{Tab:Material} in Methods). To create the data set where all parameters vary simultaneously, for the outline we used a {\em capped} Gaussian distribution of variations with a 5\% spread; for the thickness and the parameters of the material we used a {\em non-capped} Gaussian variation with 10\% spread. We adopted different distributions because the 
outline tends to be a great deal more sensitive to variations of the 20 geometric shape parameters
whereas sensitivity towards parameters describing material and thickness is much more limited. 

%*** I am not sure I understand what you wrote in the following sentences commenting Fig. 7a, and consequently I probably messed things up there. Please check. What I see is that outline and material seem very similar in 7a, while thickness has half the spread...

\begin{figure}[h]
    \centering
    \includegraphics[height=2.in]{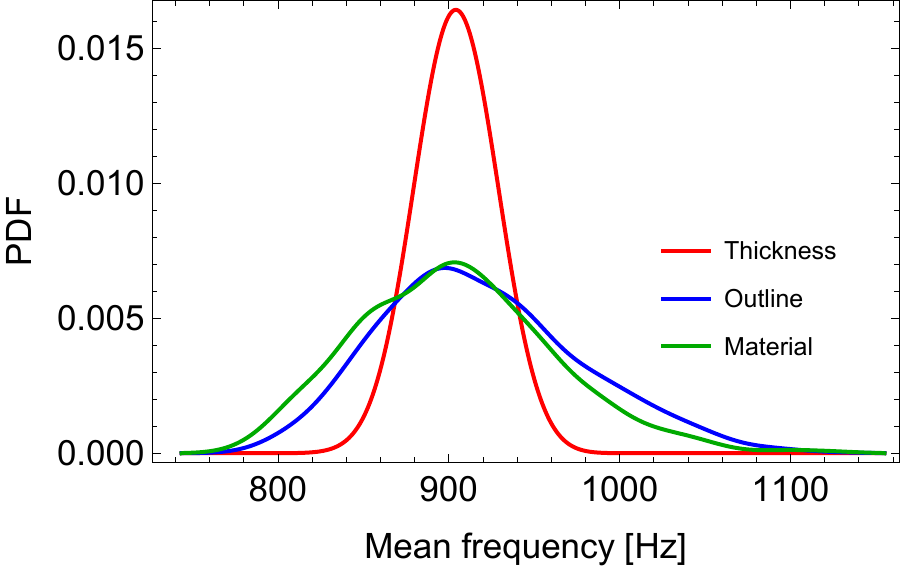}
     \put(-225,145) { \bf a)}
     \put(10,145) { \bf b)}
     \put(-225,-5) { \bf c)}
    \hspace{.25in}
    \includegraphics[height=2.in]{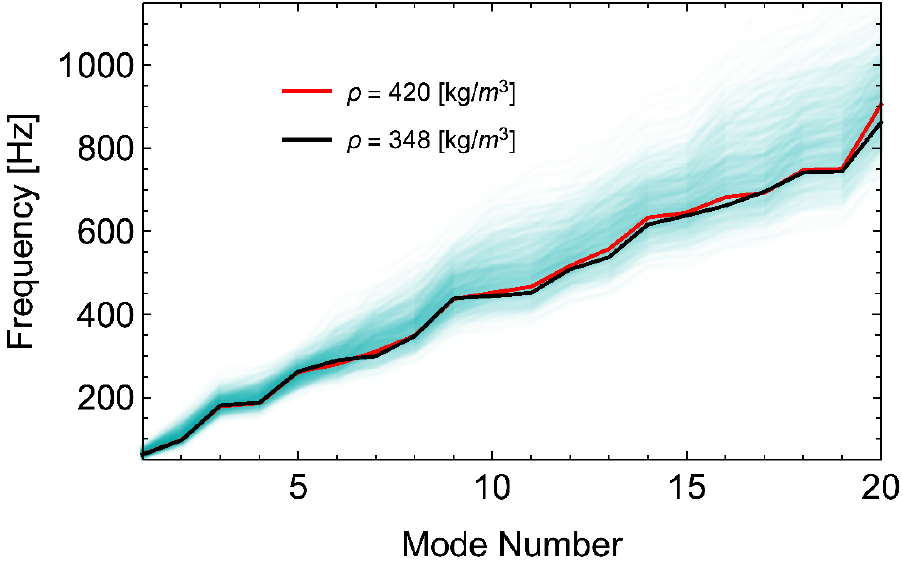}\\
    \includegraphics[height=2.3in]{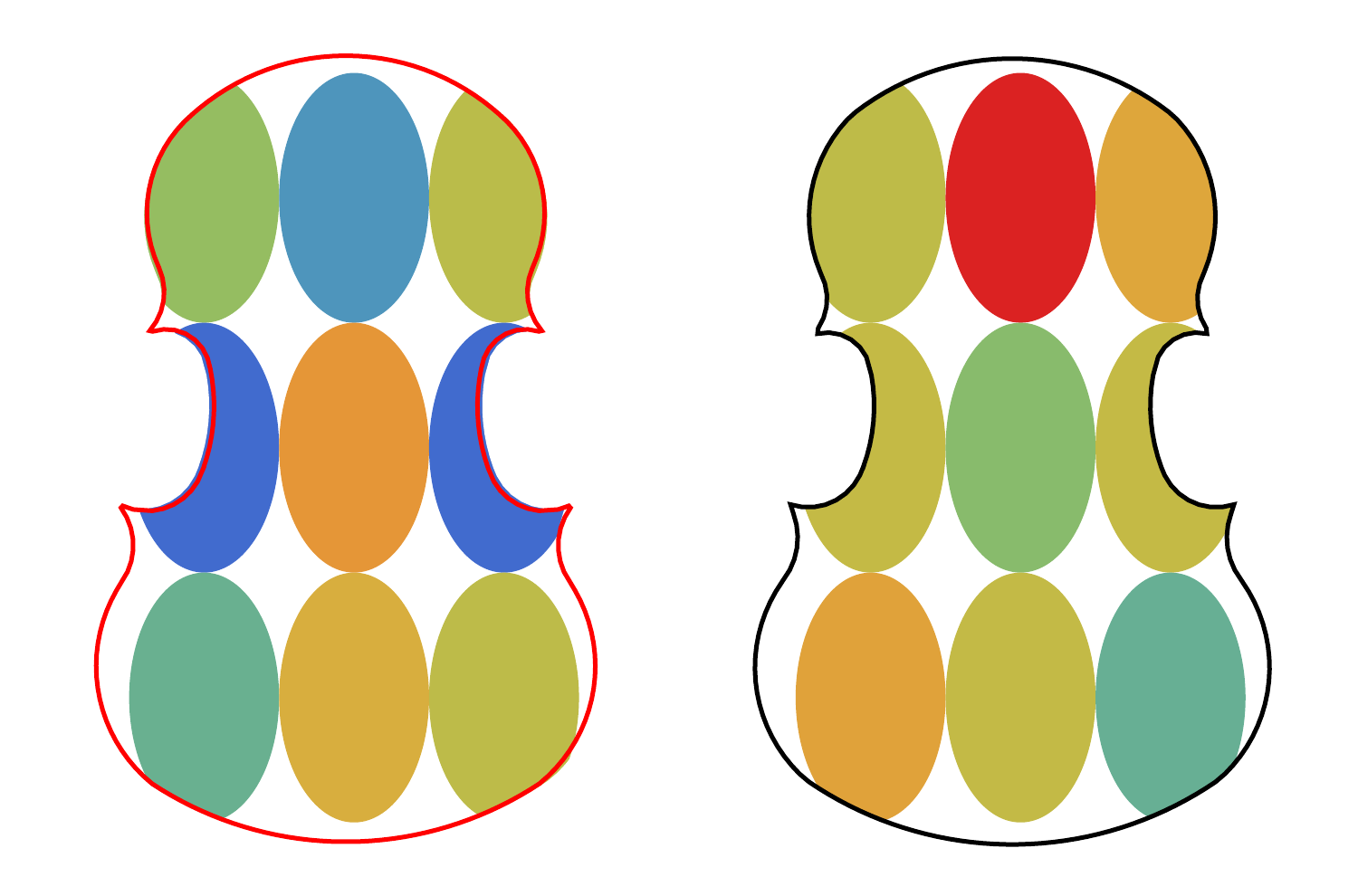}
    \put(-330,110) { \color{red}  $E_y = 1.28*10^{10}  ~$[GPa]}
    \put(-330,95) { \color{red} $\rho = 420~$[kg/$m^3$]}
    \put(-330,80) { \color{red} $v_1 = 5534~$[m/s]}
    \put(-330,55) {  $E_y = 8.99*10^{9} ~$[GPa]}
    \put(-330,40) {  $\rho = 348 ~$[kg/$m^3$]}
    \put(-330,25) {  $v_2 = 5075 ~$[m/s]}
    \includegraphics[height=2.1in]{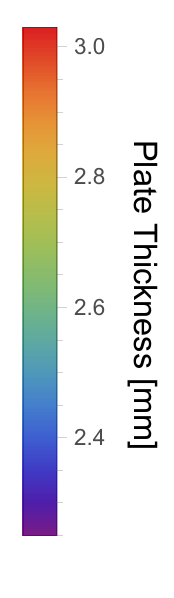}

    \caption{a) Histogram of mean frequencies for datasets varying shape, thickness profile and material parameters. 
    b) First 20 eigenfrequencies for the two most different violin top plates (in their parameters) that are in the top $0.1\%$ of most similar frequency response in the data set (the complete data set is plotted on the background for comparison). c) Outline and thickness profile of the same two violin tops. and their most important material parameters. The black outline is slightly wider than the red one and the lower corners point more upwards; the most relevant difference is in the thickness profile though.}
    \label{fig.7}
\end{figure}
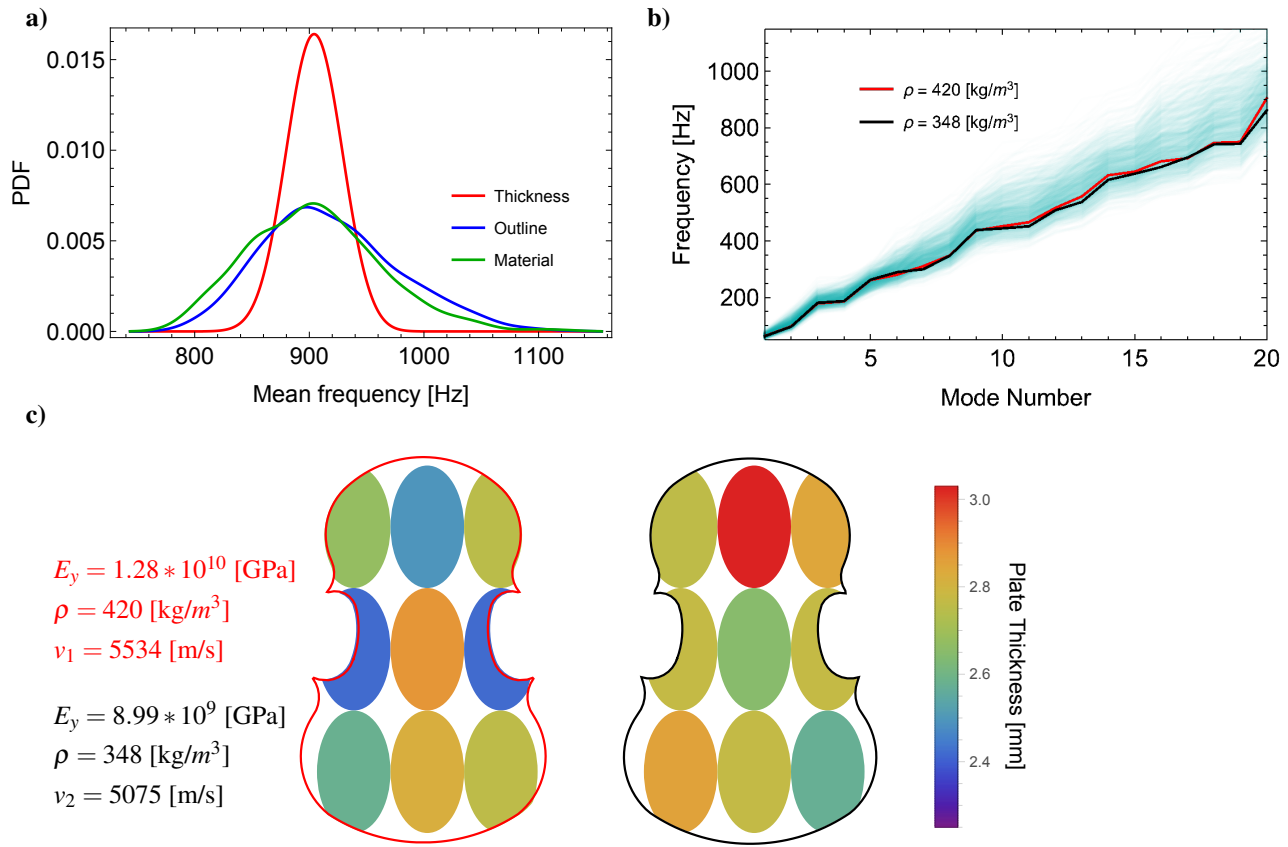

Figure \ref{fig.7}(a) shows the smoothed histogram of the average of the modal frequencies ($f_{1,...,5}$) for data sets in which each set of parameters vary independently. As we can see, the outline and the parameters of the material exhibit the same spread of mean frequency, and similar results are obtained for individual frequencies. Notice also that the distribution of the
%*** this is what I don't understand:
geometric shape parameters %*** what are the shape parameters?
has a spread that is half of that of the parameters of the material, yet the frequency spread is the same.
The histogram relative to thickness variations, on the other hand, turns out to be more {\em peaky}, which suggests that the impact of plate tuning is not as relevant as that of changing the outline and/or the properties of the material.

Finally, we wanted to understand whether plates that exhibit matching vibrational behaviour are, in fact, also similar in geometry and thickness. We picked the two violin top plates whose eigenfrequencies are the closest ones in the dataset, as shown in  Fig.~\ref{fig.7}(b); and plotted their outline and thickness profile, as shown in Fig.~\ref{fig.7}(c). We immediately notice how different the stiffness and the density of the material are (though longitudinal sound speeds are comparable $v_1 = 5534~$[m/s] and $v_2 = 5075~$[m/s]), and how such differences appear to be compensated by a balanced mix of outline and thickness variations. The top plate on the left-hand side of Fig.~\ref{fig.7}(c) (red outline) has a denser material, and its outline is thinner and narrower. The one on the right-hand side  (black outline) has a much less dense material, but and has a thicker waist. Interestingly enough, the thickness profile turns out to be asymmetrical in both cases. 
In principle, this study suggests that it is possible to create an acoustic copy of a historical instrument (same vibrational response) using material that has extremely different properties, through a careful selection of outline and thickness profiles, and it also tells us how to achieve this purpose in practice.
We believe this could have significant implications on the practice of violin making, which for the past two centuries has primarily relied on matching the shape of historical instruments. Instead, we just saw how intimately related shape and material are and, most of all, we saw how artificial intelligence can be put to work, and used to {\em design the outline} and the thickness profile in such a way to compensate for variations in material, or explore material with different properties. 

\section*{Discussion}
What we presented above is a parametric representation of the outline and of the arching of violin top plates, which allowed us to synthetically generate a rich database of geometries. We derived through simulation the eigenfrequencies of such top plates, and trained neural networks in order to be able to speed up the prediction of eigenfrequencies of nearly three orders of magnitude (600 times faster) with respect to FEM simulation, using a simple interpreted coding language such as Matlab\texttrademark.

This is, to the best of our knowledge, the first time a method is proposed for computing the geometry that can deliver the desired vibrational response in violin top plates, given the properties of the material. We showed, in fact, that neural networks can accurately predict the frequency distribution from a limited set of values. There is a clear correlation between the geometry and the eigenfrequencies, which the network can easily infer. The NN approach that we used proved to work with the thickness profile of the plate and the parameters of its material, and there is nothing to suggest that it could not be used to learn the influence of the arching profile as well, which is one important variable of the violin design that was not contemplated in this study. We are certain that this method can be easily applied to other domains of FEM simulation and maybe used to speed up the computation with simple geometries in current FEM solvers. Furthermore, recent experiments in learning the modal response of plates seem to indicate that this approach could also be used to predict the acoustic directional response of the modes as well as their frequency \cite{olivieri2020nahcnn}.

A number of luthier-specific conclusions can be drawn from study. The first one is that there is no way to independently control eigenfrequencies using either the shape of the outline or the thickness profile, as the related parameters are all intimately intertwined. Perhaps the only exception is the mode frequency $f_5$, which seems to be mostly affected by specific parameters of the outline. 
The second conclusion is that $f_5$ is mostly dependent on the `shape' of the lower bout (on the thickness of the sides or, in the case of constant thickness, on the width). The third conclusion we want to draw is that the nine regions that can be seen in historical examples of the thickness profile are quite reasonable: the correlation between the modal frequencies is symmetric and depends either on the sides or on the central region for upper bout, waist or lower bout. 
A fourth conclusion is that, assigning a NN the task of optimizing the violin design is easier, faster and less expensive: using a NN only requires the input of the parameters to be used. The final, and probably the most important conclusion of this study, is the fact that variations in the material parameters\cite{viala2020simultaneous} can only be compensated by changes in the outline of the violin, which raises doubts on how sound the contemporary practice of making copies of historical violin is, when based solely on geometry. If the material parameters are not the same, there is no hope that the instrument will vibrate (and sound) the same as the historical instrument whose geometry it matches. Further studies of the influence of the arching and the dependency of the acoustic response on parameter variation are needed to understand how to properly reproduce or, even better, how to improve on the sound of the ``old masters". 

% Despite our results consider only simulations, we have given a bold step towards a complete understanding of all the parameters affecting violin making. We have shown the correlations between geometry and eigenfrequencies, in both qualitative and quantitative terms. Instead of trying to homogenise as much as possible the material with which realising the experiments, our predictive approach could be used to study cases where material variations are compensated by geometrical changes, reducing the variability of the experimental samples in the vibrational domain. This is one of the greatest difficulties that current violin making experiments have. Furthermore, it would allow for the creation of vibrational copies of historical instruments despite not being able to match material properties of old wood, one of the biggest technical difficulties in copying old violins. 

In this article we focused on the prediction of the eigenfrequencies of a free-plate from geometry and material. We believe, however, that it will be possible to generalise and apply the same methodology to the model of a complete violin, with a focus on both its vibrational and acoustic response, which will allow us to effectively predict how a certain violin shape will `sound' given the properties of its material. The ability to predict how a violin design sounds, can truly be a game changer for violin makers, as not only will it help them do better than the ``grand masters", but it will also help them explore the potential of new designs and materials. This research allowed us to take the first steps on this path, showing how artificial intelligence, physical simulation and craftsmanship can all come together to shed light on the art of violin making. 

\section*{Methods}\label{methods}

\subsection*{FEM simulations}

The material used for the simulations is Sitka Spruce, not a common material in European violin making tradition, but with mechanical properties similar known and similar to tonewood \cite{woodhandbook}. The same wood is used also for the bass bar but, since is it not aligned to the plate, we have to pay attention to its grain orientation rotating it to the same angle of rotation of the bar.

Since the wood is an orthotropic material, we have to define different properties for the different directions. The considered values are taken according to \cite{woodhandbook} and can be seen in Table~\ref{Tab:Material}, while the density is equal to $\rho = 450 kg/m^3$. 

For the varying material parameters simulations we vary:

\begin{equation}
    \rho = \rho_0(1+\delta_\rho)\\
    E_y = E_y^0(1+\delta_{E_y})\\
    E_x = E_x^0(1+\delta_{E_x})\\
    E_z = E_z^0(1+\delta_{E_z}),
\end{equation}
where $E_i$ is the Young modulus along each dimension. The shear moduli are dependent on those: for Sitka spruce    
\begin{equation}
    G_{yx} = 0.061E_y\\
    G_{xz} = 0.064E_y\\
    G_{yz} = 0.003E_y,
\end{equation}

The same holds for the Poisson's ratios:
\begin{equation}
    \mu_{yx} = 0.467(1+\delta_{\mu_{yx}})\\
    \mu_{xz} =0.372(1+\delta_{\mu_{xx}})\\
    \mu_{yz} =0.435(1+\delta_{\mu_{yz}}),
\end{equation}
Each of the $\delta$ is independently drawn from a Gaussian distribution of zero mean and variance of 10\%, so density, stiffness and poison ratio vary independently whereas the shear moduli varies only due to the variation in $E_y$. 

The mechanical behaviour is studied through a FEM simulation performed with Comsol Multiphysics\textregistered ~software, performing an eigenfrequency study in solid mechanics physics. Each generated mesh is imported and analysed in free boundary conditions with a tetrahedron mesh automatically generated by the software.

\begin{table}[h]
\centering
\footnotesize
\begin{tabular}{|c|c|c|}
\hline
Young's Modulus & Rigidity Modulus & Poisson's Ratio \\ \hline
$E_y = 10.8$ [GPa] & $G_{yx}/E_y$ = 0.061 & $\mu_{yx}$ = 0.467 \\ \hline
$E_x/E_y$ = 0.043 & $G_{xz}/E_y$ = 0.064 & $\mu_{xz}$ = 0.372      \\ \hline
$E_z/E_y$ = 0.078 & $G_{yz}/E_y$ = 0.003 & $\mu_{yz}$ = 0.435  \\ \hline
\end{tabular}
\caption{Values of the orthotropic properties of the simulated material, density $\rho = 450kg/m^3$.}
\label{Tab:Material}
\end{table}

\subsection*{Architecture and training the neural network}
We used a feed-forward neural network, with a single hidden layer and a sigmoid activation function connected to a linear output layer. The complete structure can be seen in Figure \ref{fig.predict}.
The fully connected structure is fed with the 20 parameters that determine the shape of the outline and returns as output the eigenfrequency values of the first 10 vibrational modes of the plate. The training was done on different parts of the dataset and the evaluation of the quality in the test set, which was not seen during training.  

\subsection*{Arching and thickness}
The arching of a violin (the curvature of the top plate) is not independent of the outline. Violin makers talk about `Stradivarius' or `Guarneri' models to refer to both the outline and the curvature of the plate (the former being generally with a larger outline and higher and rather flat arching, while the latter being typically smaller and with a lower arching, albeit rounder). In this study, we intentionally chose to disregard this particular dependency. We used, in fact, one parametric shape for the arching, approximating an actual historical violin, and explored the corresponding parametric ``shape space" of outlines.

In our case, the longitudinal arching (running from neck to tailpiece of the violin) is fitted in a similar way as in \cite{gonzalez2020a}, taking the Stradivarius `Messiah' as reference for our analysis. The Messiah was on loan in 2017 at the Cremona Violin Museum for a few months, where we had access to it for analysis. 
Starting from that, we approximated its longitudinal arching ($y$ direction) using a polynomial of order 6, which is the lowest order that allows us to obtain reliable results, Fig.~\ref{fig.circles} c). The transversal arching (perpendicular to the longitudinal direction at the coordinate $y$), is approximated by a 4th-order even polynomial whose value and derivatives are zero at the edges, i.e.
\begin{equation}
a(y) + b x^2 + cx^4 = 0 \quad \textrm{and}  \quad 2bx + 4cx^3 = 0,
\label{cross_arching_system}
\end{equation}
where $a(y)$ is the elevation of the plate at the center of the plate, origin of the reference frame (see Fig.~\ref{fig.circles} d), and the values of the coefficients $b$ and $c$ are derived for each value of $y$ and $a(y)$ from the constraints in \eqref{cross_arching_system}. We assumed left-right symmetry for this model, though this assumption could be easily removed by adding odd terms in the equation.

Finally, all the violin tops are built so that they have either the same constant thickness of $2.7$ mm in the arched region (when we vary only the outline) or varying thickness in 9 different regions as in Fig.~\ref{fig.thicknessCorr}. The actual procedure for the creation of the top meshes can be found in \cite{gonzalez2020a}. We have chosen the $2.7$mm thickness as representative of violins based on the historical results found in \cite{stoel2008comparison}.

\subsection*{Dataset creation}
The dataset is created starting from the outline that best fits the Messiah (area difference $<1\%$). We found the parameters using an iterative numerical optimization method. As the area difference between two violin outlines is an extremely nonlinear function of the shape parameters, simple minimization procedures are not readily applicable. From that outline we randomly modify each of its parameters $p_i$ using a Gaussian distribution ($P(x) = \frac{1}{\sigma \sqrt{2\pi} } e^{-\frac{1}{2}\left(\frac{x}{\sigma}\right)^2}$) of zero mean and a standard deviation of 5\% as follows
%$\sigma =0.05$ 
\begin{equation}
    p_i = p_{i}^0 (1 + \delta_i)~,
\end{equation}
where each $\delta_i$ is independently and normally distributed but capped at 0.1, so the maximum variation in each parameter is a 10\%. This cap is necessary to prevent top plates that come from the Gaussian distribution of the parameters from drifting too far apart from ``standard" shapes. In fact, even a 15\% cap would easily originate strange shapes. 

The values that we chose generated a rather diversified variety of violin shapes, most of which are similar to historical examples, though some of them would resemble a \emph{Viola da Gamba} or a very skinny fiddle. The resulting dataset consists of 1750 top plates of varying outline, 1000 top plates of fixed outline and varying thickness, 1000 top plates of varying material parameters and 1500 of varying thickness, outline and material parameters. All the violin tops in the data set exhibit the same longitudinal arching but the transverse ones are different, though each one of them functionally equivalent and obeying the same boundary conditions \eqref{cross_arching_system} on its own outline.

\section*{Supplementary material 1}

Equation defining the 17 points and 20 parameters used. $p_i$ stands for ${x,y}$ points whereas alphabetic letters are one dimensional parameters. This will be given as a Mathematica/Matlab code the reader can run and create different violin plates. 

\begin{verbatim}

p0 = {-x0, 0};
p1 = p0 + aa {Cos[a], Sin[a]};
p2 = p0 + aa {Cos[b], Sin[b]};
p3 = p2 - c {Cos[b], Sin[b]};
p4 = p1 - cc {Cos[a], Sin[a]};
p5 = p4 + cc {Cos[d], Sin[d]};
p6 = p3 + c {Cos[e], Sin[e]};
p7 = p6 + f {Cos[g], Sin[g]};
p8 =  p5 + h {Cos[d2], Sin[d2]};
p9 = p8 + h {Cos[l], Sin[l]};
p10 = p9 + k {Cos[l], Sin[l]};
p11 = p10 +  k {Cos[\[Pi]/2 + hh], Sin[\[Pi]/2 + hh]};
p12 = p11 +  k {Cos[hh], Sin[hh]};
p13 = p7 + f {Cos[gg], Sin[gg]};
p14 = p13 + ff {Cos[gg], Sin[gg]};
p15 = p14 + ff {Cos[-\[Pi]/2 - kk], Sin[-\[Pi]/2 - kk]};
p16 = {0, p15[[2]] + rr2 Cos[ArcSin[Sin[kk] ff/rr2 - p14[[1]]/rr2]]};
p17 = {0, p11[[2]] - rr Cos[ArcSin[Sin[hh] k/rr - p10[[1]]/rr]]};\end{verbatim}

\bibliography{bib.bib}

\section*{Acknowledgements}

We would like to thank the Ashmolean Museum in Oxford, UK, for giving us access to the Stradivarius ``Messiah" and  authorising the 3D scanning of the instrument. We are grateful to the Violin Museum Foundation in Cremona for giving us access to the Stradivarius ``Messiah" while it was exhibited there. We particularly grateful to Colin Harrison, Senior Curator of European Art in the Department of Western Art in the Ashmolean Museum; and the prominent contemporary American violin maker Gregg Alf, for making the measurement session of the instrument possible in the Musical Acoustics Labs of the Politecnico di Milano, which are located at the Violin Museum of Cremona, Italy. We also grateful to the `Arvedi Non-Invasive Diagnostic Lab' from the University of Pavia, with whom we share the premises of the scientific Labs as well as the data that we collect together. This work has been funded by the Cultural District of Violin Making of the City of Cremona, Italy.

\section*{Author contributions statement}
S.G. conceived the experiments, D.S. conducted the experiments, S.G. and D.B. performed the PCA/correlation analysis. All authors analysed the results and reviewed the manuscript. 

\section*{Additional information}

To include, in this order: \textbf{Accession codes} (where applicable); \textbf{Competing interests} (mandatory statement). 

The corresponding author is responsible for submitting a \href{http://www.nature.com/srep/policies/index.html#competing}{competing interests statement} on behalf of all authors of the paper. This statement must be included in the submitted article file.

\end{document}